\documentclass[11pt,A4paper,conference]{IEEEtran}
\usepackage{amsmath,amssymb,eucal,graphicx,times,exscale,epsfig}
\usepackage[usenames, dvipsnames]{color}
\usepackage{rotating}
\usepackage{lscape}
\usepackage{bezier}
\newtheorem{theorem}{Theorem}

\newtheorem{lemma*}[theorem]{Lemma}

\def\L{\mathcal{L}}

\def\R{\mathcal{R}}

\def\L{\mathcal{L}}

\def\awgn{\ifmmode\mathrm{BIAWGN}\else BIAWGN\fi}

\def\phi{\varphi}

\begin{document}

\title{On Properties of the Minimum Entropy Sub-tree\\ to Compute Lower Bounds on the Partition Function}

\author{\authorblockN{Mehdi Molkaraie}
\authorblockA{ALGO, IC, EPFL \\ %Ecole Polytechnique F\'ed\'erale de Lausanne \\
%Ecole Polytechnique F\'ed\'erale de Lausanne \\
CH-1015 Lausanne \\
Switzerland \\
{\bf {\small mehdi.molkaraie@a3.epfl.ch}}}
\and
\authorblockN{Payam Pakzad}
\authorblockA{Digital Fountain\\
Fremont, CA 94538\\
USA \\
{\bf {\small payam@digitalfountain.com}}}
}
%\date{(Extended Abstract)}
\maketitle
\begin{abstract}  
Computing the partition function and the marginals of a global probability 
distribution are two important issues in any probabilistic inference problem. 
In a previous work, we presented sub-tree based upper and lower 
bounds on the partition function of a given probabilistic inference 
problem. Using the entropies of the sub-trees we proved an inequality 
that compares the lower bounds obtained from different sub-trees. In this paper 
we investigate the properties of one specific lower bound, namely the lower 
bound computed by the minimum entropy sub-tree. We also investigate the relationship 
between the minimum entropy sub-tree and the sub-tree that gives the best 
lower bound.

%Finally we propose a greedy algorithm that computes 
%low-complexity bounds on the partition function. 
%Simulation results on two-dimensional grids are reported.
%In the final version of this document, we will also report simulation results on the quality of
%our bounds in practical settings.
\end{abstract}
\section{Introduction}
\label{se:intro}
The {\it partition function} is of great importance in statistical physics since most of the
thermodynamic variables of a system can be expressed in terms of this quantity or its
derivatives. This quantity also plays an important role in many other contexts, including 
artificial intelligence, combinatorial enumeration, approximate inference, and parameter estimation. 
In general, the exact calculation of the partition function is computationally intractable 
therefore finding low-complexity estimates and bounds is desirable.

In~\cite{MoPa:05}, we proposed upper and lower bounds on the partition function that 
depend on the partition function
of any sub-junction tree of a given junction graph representing the inference problem. 
In~\cite{MoPa:06} a greedy algorithm that gives low-complexity upper and lower bounds 
on the partition function was proposed. An inequality was proved
that compares the lower bounds calculated from different sub-junction trees based on their entropies
~\cite[Theorem 2]{MoPa:06}.

In this paper, we study the properties of the minimum entropy sub-junction tree and 
will extend the results of~\cite[Theorem 2]{MoPa:06} by stating new theorems and corollaries. We prove that there is an upper bound
on how much any other lower bound can be better than the one 
obtained from the minimum entropy sub-tree.
We also show that the probability distributions over the sub-tree that gives 
the best lower bound and the one with the minimum entropy are close in divergence .
%~\ref{th:Twotrees}, which suggests a method to compare bounds obtained from different sub-trees. 
%In section~\ref{se:Z} we propose a greedy algorithm that finds a maximal tree;
%a maximal tree gives bounds on the partition function that are at least as good as the bounds 
%obtained from all its sub-trees. 
%Simulation results on two-dimensional grids with different sizes and edge strengths are 
%reported in section~\ref{se:Simul}. 
%\begin{figure}[t]
%\begin{center}
%\psfig{file=JGraph2.eps,width=41mm}\hspace{5mm}
%\psfig{file=JGraph22.eps,width=41mm}
%\end{center}
%\caption{Two junction graphs representing the same inference problem}
%\label{fig:JGexample1}
%\end{figure} 

\section{Background}
\label{se:inference}
Suppose a global function defined over several random variables, e.g. a probability 
mass function, factors as a product of a series of non-negative local kernels,
each kernel defined over a subset of the set of all random variables. The goal is to 
compute the normalization constant and the marginals of the global function according
to those subsets.

More formally, consider a set $\{X_1,X_2,\ldots,X_N\}$ 
of $N$ discrete random variables taking their values in a finite 
set $A=\{0, 1, 2,\ldots, a-1\}$. Let $x_i$ represent the possible realizations 
of $X_i$ and let ${\bf x}$ stand for $\{x_1,x_2,x_3,\ldots,x_N\}$. Suppose
$R_1,R_2,\ldots,R_M$ are subsets of $\{1,2,\ldots,N\}$ and
$\mathcal{R}=\{R_1,R_2,\ldots,R_M\}$ is a collection of subsets of the indices 
of the random variables $X_1$ through $X_N$. Let us also suppose that 
$p({\bf x})$, the joint probability mass function, factors into product of finite and
non-negative local kernels as
\begin{eqnarray}
\label{eq:jointp}
  p({\bf x}) = \frac{1}{Z}\prod_{R\in \mathcal{R}}\alpha_R({\bf x}_R),
\end{eqnarray}
where each local kernel $\alpha_R(x_R)$ is a function of the variables whose indices appear in $R$, and $Z$
is the partition function, also known as the {\em global normalization constant} whose role
is simply normalizing the probability distribution.

In a {\it probabilistic inference} problem, we are interested in computing $Z$ and the 
marginal densities $p_R({\bf x}_R)$, which are defined as
\begin{eqnarray}
\label{eq:ZZ}
%\begin{array}{rcl}
  Z & = & \displaystyle\sum_{\bf x}\prod_{R\in \mathcal{R}}\alpha_R({\bf x}_R),\\
  p_R({\bf x}_R) & = & \displaystyle\sum_{{\bf x \setminus x}_R} p({\bf x}).
%\end{array}
\end{eqnarray}

\section{Graphical Models and the Generalized Distributive Law}
\label{se:gdl}
Graphical models use graphs to represent and manipulate joint probability distributions.
An efficient way to solve a probabilistic inference problem is to represent it with a 
graphical model and use a message passing algorithm on this model. 

There are many graphical models in the literature such as junction graphs, Markov 
random fields, and (Forney-style)
factor graphs. In this paper we focus on graphical models defined in terms of {\it junction graphs}.
Our results can be easily expressed with other graphical models.
\vspace{0.08cm}
%\section{Junction Graphs}
%\label{se:junction}

{\it Definition 1:} A junction graph is  an undirected graph $\mathcal{G}=(V,E,L)$ where each
vertex and each edge have labels, denoted by $L(v)$, and $L(e)$ respectively. The labels 
on the edges must be a subset of the labels of their corresponding vertices.
Furthermore, the induced subgraph consisting only of the vertices and edges which contain 
a particular label, must be a tree \footnote {There is a generalization for this definition known as 
%the generalized junction graph and a furthur generalization known as 
region graphs,see~\cite{YFW:05}; for simplicity we prefer
to work with junction graphs.}.

\vspace{0.09cm}

We say that $\mathcal{G}=(V,E,L)$ is a junction graph for the inference problem
defined by $\mathcal{R}$, 
if $\{L(v_1), L(v_2), \ldots, L(v_M)\}=\mathcal{R}$. For any probabilistic inference 
a junction graph representation always exists. 

%In~\cite[Theorem 4]{AM:01} it is proved that for any probabilistic inference problem there is 
%always a junction graph representation (but not necessarily a junction tree representation). 
%
%In general there might be many junction graphs representing the same inference problem. 
%For example Fig.~\ref{fig:JGexample1} shows two junction graphs for 
%$\mathcal{R}=\{\{1,2,5\},\{2,3,5\},\{3,4,5\},\{1,4,5\}\}$.

{\it The generalized distributive law (GDL)} is 
an iterative message passing algorithm, described by its \emph{messages} and \emph{beliefs}, 
to solve the probabilistic inference problem on a junction graph. It operates by passing
messages along the edges of a junction graph, see \cite{AM:00}, \cite{SS:90}.

The  message sent from a vertex $u$, to another vertex $v$, is a function 
of the variables whose indices are on $e$, the edge between $v$ and $u$, and is 
denoted by $m_{u,v}({\bf x}_{L(u,v)})$. The beliefs on vertices and edges are denoted 
by $b_v({\bf x}_{L(v)})$ and $b_e({\bf x}_{L(e)})$, respectively. The messages and the beliefs 
are computed as
\begin{multline*}
m_{u,v}({\bf x}_{L(u,v)})  =\\  \displaystyle\sum_{{\bf x}_{L(u)\setminus L(v,u)}} 
                             \alpha_{u}({\bf x}_{L(u)})
                             \prod_{u' \in N(u)\setminus v} m_{u',u}({\bf x}_{L(u',u)}).
\end{multline*}

\begin{eqnarray*}
  b_v({\bf x}_{L(v)}) & = & \frac{1}{Z_v}\alpha_{v}({\bf x}_{L(v)})\displaystyle\prod_{u\in N(v)}
                             m_{u,v}({\bf x}_{L(u,v)}) \\
  b_e({\bf x}_{L(e)}) & = & \frac{1}{Z_e}m_{u,v}({\bf x}_{L(e)})m_{v,u}({\bf x}_{L(e)}),
\end{eqnarray*}
where $N(v)$ denotes the neighbors of $v$; $Z_v$ and $Z_e$ are the local normalizing constants.

\vspace{0.10cm}

\begin{theorem}
\label{th:GDL}
On a junction tree the beliefs converge to the exact local
marginal probabilities after a finite number of steps~\cite[Theorem 3.1]{AM:00}.
\end{theorem}

%\vspace{0.11cm}

If $\mathcal{G}$ is a tree, $p({\bf x})$ defined by (\ref{eq:jointp}) factors as follows, see~\cite{CO:98}
\begin{equation}
\label{eq:decomp}
   p({\bf x})=\frac{\prod_{v \in V}{p_v({\bf x}_{L(v)})}}{\prod_{e \in E}{p_e({\bf x}_{L(e)})}}.
\end{equation} 

In this case, the entropy of the global distribution decomposes as the sum of the 
entropies of the vertices minus the sum of the entropies on the edges.

Similarly, the global normalization constant $Z$ can be expressed in terms of the local
normalization constants as follows
\begin{equation}
\label{eq:decompZ}
   Z =\frac{\prod_{v \in V}{Z_v}}{\prod_{e \in E}{Z_e}}.
\end{equation} 

%\vspace{0.12cm}

Therefore if $\mathcal{G}$ is a tree, there is an efficient algorithm to compute $Z$, the marginals 
of $p({\bf x})$, and the entropy of $p({\bf x})$. If $\mathcal{G}$ is not a tree, the 
above algorithm is not guaranteed to give the exact solution or even to converge, although
empirically it performs very well. 

%This is the topic of the next section.
%show that the fixed points of the algorithm correspond to the stationary points of
%a certain approximation to some energy function. This is the topic of the next section.
%
\section{Connection to Statistical Physics}
\label{se:Stat} 
New theoretical results show that there is a 
connection between message passing algorithms and certain approximations to 
the energy function in statistical mechanics. The idea is that having plausible approximations
to the energy function gives hope that the minimizing arguments are also reasonable approximations
to the exact marginals, see~\cite{YFW:05, PA:02}. 
See also~\cite{CC:06} for some new results regarding the partition function and loop series. 

\section{Sub-tree Based Lower Bounds on the Partition Function}
\label{se:Bounding}

For a general junction graph, calculating the partition function, $Z$, through a 
straightforward manner as expressed in (\ref{eq:ZZ}), needs a sum with an exponential 
number of terms. 
Therefore it is desirable to have
bounds on $Z$ which can be obtained with 
low complexity, see~\cite{WJW:02}.
 
According to (\ref{eq:decompZ}), on a junction tree the partition function can be computed 
efficiently. In this section we derive lower bounds on $Z$ which 
depend on the partition function 
of $\mathcal{G}_T$ a sub-junction tree of $\mathcal{G}$, see~\cite{MoPa:05, MoPa:06}. 

Consider
a probabilistic inference problem defined by $\mathcal{R}=\{R_1,R_2,R_3,\ldots,R_M\}$. Also consider
$\mathcal{R}_T$, a subset of $\mathcal{R}$ that has a junction tree representation. If $q_{T}({\bf x})$ denotes 
the global probability distribution and $Z_T$ the partition function
constant on $\mathcal{G}_T$, we can 
rewrite $p({\bf x})$ defined in (\ref{eq:jointp}) as follows

\begin{align}
\label{eq:rewrite}
  p({\bf x}) & = \frac{1}{Z}\prod_{R\in \mathcal{R}_T}\!\!\alpha_R({\bf x}_R)
               \!\!\!\prod_{R\in \mathcal{R} \setminus \mathcal{R}_T}\!\!\!\!\!\alpha_R({\bf x}_R)\notag\\
             & = \frac{Z_T}{Z}\, q_T({\bf x})\!\!\!\!\!\prod_{R\in \mathcal{R} \setminus \mathcal{R}_T}\!\!\!\!\alpha_R({\bf x}_R).
\end{align}

%\subsection{Upper Bound}
%To derive the upper bound, we consider $D\big(p({\bf x})||q_T({\bf x})\big)$. Using (\ref{eq:rewrite}) we can write
%\begin{multline}
%\label{eq:UB1}
%  D\big(p({\bf x})||q_T({\bf x})\big)\;
%             = \;D\big(p({\bf x})||\frac{Z}{Z_T}\cdot\frac{p({\bf x})}{\prod_{R\in \mathcal{R} \setminus \mathcal{R}_T}\!\!\!\!\alpha_R({\bf x}_R)}\big)\\ = \ln(\frac{Z_T}{Z})
%               +\sum_{R\in \mathcal{R}\setminus \mathcal{R}_T} \sum_{{\bf x}_R}  
%               p({\bf x}_R)\ln\alpha_R({\bf x}_R)
%\end{multline}
%
%Using the non-negativity of $D\big(p({\bf x})||q_T({\bf x})\big)$ we have
%\begin{equation}
%\label{eq:UB3}
%  \ln(Z) \leq \ln(Z_T)
%  +\!\!\!\sum_{R\in \mathcal{R}\setminus \mathcal{R}_T} \!\!\sum_{{\bf x}_R}  
%               p({\bf x}_R)\ln\alpha_R({\bf x}_R)
%\end{equation}

%\subsection{Lower Bound}
Take logarithm of both sides of (\ref{eq:rewrite}), multiply by $q_T({\bf x})$, and 
sum over ${\bf x}$.
\begin{multline}
\label{eq:UB4}
  \sum_{{\bf x}}q_T({\bf x})\ln p({\bf x})\;
             = \;\ln(\frac{Z_T}{Z}) + \sum_{{\bf x}}q_T({\bf x})\ln q_T({\bf x})\\
               \;\;\;\;\;\, +\sum_{R\in \mathcal{R} \setminus \mathcal{R}_T} \sum_{{\bf x}} 
               q_T({\bf x})\ln \alpha_R({\bf x}_R).
\end{multline}
By rearranging (\ref{eq:UB4}) we obtain
\begin{multline}
\label{eq:UB5}
  -D\big(q_T({\bf x})||p({\bf x})\big)\;
             = \;\ln(\frac{Z_T}{Z}) \\
	\;\;\;\;\;\, +\sum_{R\in \mathcal{R} \setminus \mathcal{R}_T} \sum_{{\bf x}} 
               q({\bf x})\ln \alpha_R({\bf x}_R).
\end{multline}
Hence the following
\begin{equation}
\label{eq:UB6}
   \sum_{R\in \mathcal{R} \setminus \mathcal{R}_T} \sum_{{\bf x}} 
               q_T({\bf x})\ln \alpha_R({\bf x}_R) + \ln(Z_T) \leq \ln(Z).
\end{equation}
%This is what is usually called the structured mean field theory.

If we denote the lower bound obtained using $q_T$, i.e. the left hand side of (\ref{eq:UB6}), by $\L_{q_T}$,
the following theorem holds. See~\cite[Theorem 2]{MoPa:06}.

%\vspace{0.08cm}

\begin{theorem}
\label{th:Twotrees}
Consider $\mathcal{R}_{1}$ and $\mathcal{R}_{2}$, subsets of $\mathcal{R}$ with junction tree 
representations. Also suppose that
$q_1({\bf x})$ and $q_2({\bf x})$ denote the global probability distributions, and $Z_{1}$ and $Z_{2}$ 
the partition functions over $\mathcal{R}_{1}$ and $\mathcal{R}_{2}$ respectively. Without loss
of generality suppose $H(q_1) \le H(q_2)$, then the following inequality holds
\begin{multline}
\label{eq:Compare}
\L _{q_2} \; \leq \; \L_{q_1} + \\
            \, \!\! \min \big(D(q_1||\overline{q}_1)\!\!-\!\!D(q_2||q_1), D(q_1||q_2)\!\!+\!D(q_1||\overline{q}_2)\big).
\end{multline}
Here $\overline{q}_1$ and $\overline{q}_2$ denote the global probability distributions
on $\mathcal{R}\setminus \mathcal{R}_{1}$ and $\mathcal{R}\setminus \mathcal{R}_{2}$ respectively. 
\end{theorem}

%\vspace{0.08cm}

%\begin{corollary}
\begin{figure}[t]
\begin{center}
\includegraphics[width=35mm]{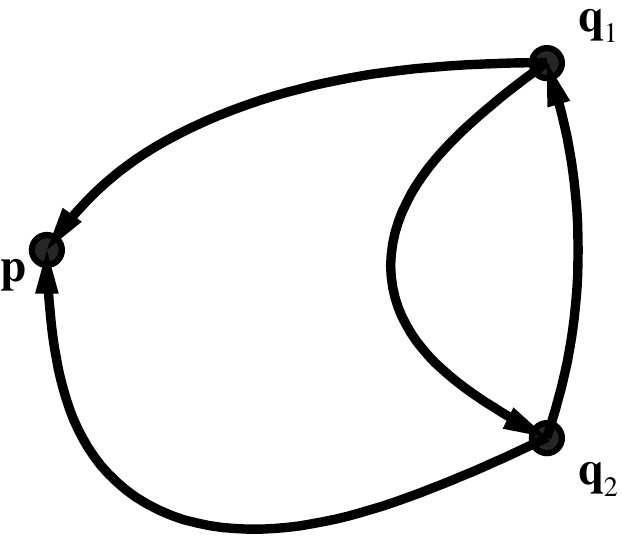}%\hspace{5mm}
\end{center}
\caption{To $p$ via $q_1$ or via $q_2$}
\label{fig:qpq}
\end{figure} 
{\it Corollary 1:}
In the case that $\mathcal{R}_{1} = \mathcal{R}\setminus \mathcal{R}_{2}$, 
namely when the junction graph decomposes into two junction trees (for example this can be the case
when we choose a sub-tree in a graph with only one cycle), if $H(q_1) \le H(q_2)$ the bound in (\ref{eq:Compare}) simplifies to
%In the case that $H(q_1) \le H(q_2)$ and $\mathcal{R}_{1} = \mathcal{R}\setminus \mathcal{R}_{2}$, 
%namely when the junction graph decomposes into two junction trees the bound 
%in (\ref{eq:Compare}) simplifies to
\begin{equation}
\label{eq:Compare2}
\L _{q_2} + D(q_2||q_1) \; \leq \; \L_{q_1} + D(q_1||{q}_2).
\end{equation}
Note that in general $D(\cdot||\cdot)$ is not symmetrical therefore 
equation (\ref{eq:Compare2}) does not tell whether one bound is better
than the other.
According to (\ref{eq:UB5}) and the definition of $\L_{q_T}$, we can see that 
$\L_{q_T} = \ln(Z)-D(q_T||p)$ therefore we can rewrite (\ref{eq:Compare2}) as
\begin{equation}
\label{eq:Compare3}
D(q_2||q_1) + D(q_1||p) \; \leq \; D(q_1||q_2) + D(q_2||p).
\end{equation}
%\end{corollary}

In other words, the distance from $q_2$ to $p$ via $q_1$ is shorter than the distance
from $q_1$ to $p$ via $q_2$. See Fig.~\ref{fig:qpq}. Note that in general 
$D(\cdot||\cdot)$ does not satisfy the triangular inequality therefore 
equation (\ref{eq:Compare3}) does not tell whether one bound is better
than the other either.

Clearly, the distance from $q_1$ to $p$ is also shorter than the distance 
from $q_1$ to $p$ via $q_2$.

\begin{equation}
\label{eq:Compare4}
D(q_1||p) \; \leq \; D(q_1||q_2) + D(q_2||p).
\end{equation}
%Note that in general $D(\cdot||\cdot)$ is not 
%symmetrical and does not satisfy the triangular inequality therefore 
%equations (\ref{eq:Compare2}) and (\ref{eq:Compare3}) do not tell 
%whether one bound is better than the other. 
%
%\vspace{0.033cm}

{\it Corollary 2:}
Consider a subset $\R_S$ of $\R$ with junction tree representation. Also suppose that 
$q_S$, the probability distribution over $R_S$, has the smallest entropy among
all the probability distributions on sub-trees. Then for any subset $\R_T$ of $\R$ with 
tree representation the following inequality holds
\begin{equation}
\L_{q_T}\, \leq \, \L_{q_S}\,+\,D(q_S||\overline{q}_S).
\end{equation} \ \\
%\end{corollary}
\begin{proof}
\vspace{-0.1cm}
According to Theorem~\ref{th:Twotrees}
\vspace{-0.1cm}
%Since $H(q_S) \leq H(q_T)$ according to Theorem~\ref{th:Twotrees} we can write. 
\begin{multline*}
\L_{q_T} \leq \L_{q_S} + \\
             \!\!\!\! \min \big(D(q_S||\overline{q}_S)\!-\!D(q_T||q_S),\! D(q_S||q_T)\!\!+\!\!D(q_S||\overline{q}_T)\big).
\end{multline*}
and hence the following
\vspace{-0.1cm}
\begin{eqnarray*}
\L_{q_T} & \leq & \L_{q_S} \!+\! D(q_S||\overline{q}_S)\!-\!D(q_T||q_S) \nonumber \\
         & \leq & \L_{q_S} \!+\! D(q_S||\overline{q}_S).
\end{eqnarray*}
\end{proof}
In other words, the lower bound obtained from any sub-tree can not be 
better than the lower bound obtained from the minimum entropy sub-tree by
more than $D(q_S||\overline{q}_S)$, a value that does not depend on $q_T$. This 
gives us a quality guarantee for the lower bound obtained from the minimum 
entropy sub-tree~\cite{MM:07}.

%\vspace{0.17cm}

\begin{theorem}
\label{th:thBS}
Consider subsets $\R_S$ and $\R_B$ of $\R$ with junction tree representations. Also
suppose that $q_S$, the probability distribution over $\R_S$, has the smallest
entropy and $q_B$, the probability distribution over $\R_B$, gives the best lower bound, then the following inequality holds
\begin{equation} 
D(q_B||q_S)\,\leq\, D(q_S||\overline{q}_S). 
\end{equation}
\end{theorem} \ \\
\begin{proof}
Since $H(q_S) \leq H(q_B)$ according to Theorem~\ref{th:Twotrees} we can write
\begin{equation}
\label{eq:th51}
\L_{q_B}\, \leq \,\L_{q_S} + D(q_S||\overline{q}_S) - D(q_B||q_S).
\end{equation}
Since $q_B$ gives the best lower bound
\begin{equation}
\label{eq:th52}
\L_{q_S} \, \leq \, \L_{q_B}.
\end{equation}
The proof would be clear by adding equations (\ref{eq:th51}) and (\ref{eq:th52}).
%For the proof add equations (\ref{eq:th51}) and (\ref{eq:th52}).
\end{proof}

%\vspace{0.12cm}

\begin{figure}[t]
\begin{center}
\includegraphics[width=37mm]{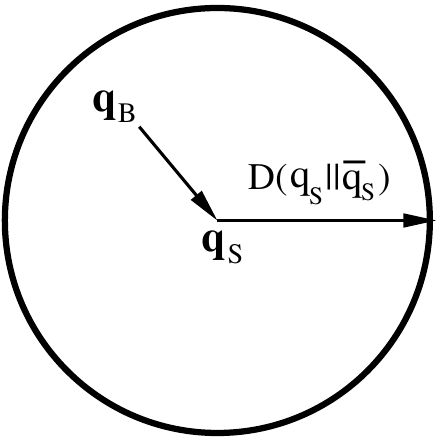}%\hspace{5mm}
\end{center}
\caption{Upper bound for divergence between $q_B$ and $q_S$}
\label{fig:Dqq}
\end{figure}

%\begin{figure}[t]
%\begin{center}
%\psfig{file=Dqq.eps,width=43mm}%\hspace{5mm}
%\end{center}
%\caption{Upper bound for divergence between $q_B$ and $q_S$}
%\label{fig:qpq}
%\end{figure} 
%
Theorem~\ref{th:thBS} gives us another quality guarantee regarding the minimum
entropy sub-tree (which is the least random, least uncertain, and most biased sub-tree). This 
theorem shows that the probability distribution on the tree that gives the best 
bound and the minimum entropy distribution are close, where closeness is measured by 
divergence. The upper bound is $D(q_S||\overline{q}_S)$ which does not depend 
on $q_B$~\cite{MM:07}. See Fig.~\ref{fig:Dqq}.

%\vspace{0.17cm}

{\it Corollary 3:}
In the case that $\mathcal{R}_{S} = \mathcal{R}\setminus \mathcal{R}_{B}$ we
have the following inequality
\begin{equation} 
D(q_B||q_S) \leq D(q_S||q_B).
%D(q_B||q_S) & \leq & D(q_S||\overline{q}_S) 
\end{equation} 

%\vspace{0.12cm}

{\it remark 1:}
In almost all the theorems and corollaries, we insisted that the subsets of $\R$
have junction-tree representations. This assumption can be relaxed and the 
theorems and corollaries would still be valid for the sub-graphs. However, 
having a junction tree representation makes the computation of the entropy and
the partition function easier (using GDL or any other iterative message passing 
algorithm).

\section{Conclusion}
In this paper, we extended some of our previous results on bounding the
partition function. In the case that the graph decomposes into two sub-trees
we derived a number of divergence inequalities concerning the global probability 
distribution and the probability distributions on the sub-trees. We showed 
that the minimum entropy sub-tree has some optimality properties, 
namely the lower bound obtained from this tree can not be far from the 
lower bound obtained from any other sub-tree and the probability distribution 
on this tree and the probability distribution on the tree that gives the
best lower bound are close where the divergence is the measure 
of closeness.

%We studied upper and lower bounds on the partition function 
%that depend on the partition functions of the sub-junction trees of a junction graph. 
%Based on the entropies of sub-trees, in a Theorem we proved an inequality 
%to compare such bounds. 
%We also proposed a greedy algorithm 
%and derived low-complexity upper and lower bounds that depend on the partition 
%function of any maximal sub-junction tree of a given junction graph.

 \section*{Acknowledgment}
The first author wishes to thank J. Dauwels and N. Macris for their helpful comments.
%\section*{References}


\begin{thebibliography}{99}

\bibitem{AM:00}S. M. Aji and R. J. McEliece. ``The Generalized Distributive Law,'' \emph{IEEE Trans. Info. Theory,} 325--343,
March 2000.

\bibitem{AM:01}S. M. Aji and R. J. McEliece. ``The Generalized Distributive Law and Free Energy Minimization,'' 
\emph{39th Allerton Conference on Communication, Control, and Computing,} October 2001.

\bibitem{CC:06}M. Chertkov and V. Chernyak. ``Loop Calculus in Statistical Physics and Information 
Science.'' Phys. Rev. E 73, June 2006.

\bibitem{CO:98}R. Cowell. Advanced Inference in Bayesian Networks. In 
\emph{Learning in Graphical Models}. MIT Press, 1998.

\bibitem{MoPa:05}M. Molkaraie and P. Pakzad. ``On Entropy Decomposition and New Bounds on the Partition Function,''
\emph{International Symposium on Info. Theory,} Australia, September 2005.

\bibitem{MoPa:06}M. Molkaraie and P. Pakzad. ``Sub-tree Based Upper and Lower Bounds on the Partition Function,''
\emph{International Symposium on Info. Theory,} Seattle, July 2006.

\bibitem{MM:07}
M.~Molkaraie, 
\emph{Subtree-Based Bounds and Simulation-Based Estimations
for the Partition Function,} Ph.D.\ dissertation $3880$, EPF Lausanne, Switzerland, 2007.


\bibitem{PA:02}P.~Pakzad and V.~Anantharam. ``Estimation and Marginalization Using Kikuchi
Approximation Methods,''
{\em Neural Computation,} 1836--1873, August 2005.

\bibitem{SS:90}G. R. Shafer and P. P. Shenoy. ``Probability Propagation,'' 
\emph{Annals of Mathematics and Artificial Intelligence,} 
327--352, 1990.

\bibitem{WJW:02}M. J. Wainwright, T. Jaakkola, and A. S. Willsky. ``A New Class of Upper Bounds on the Log Partition
Function,'' \emph{IEEE Trans. Info. Theory,} 2313--2335, July 2005.

\bibitem{YFW:05}J. S. Yedidia, W. T. Freeman, and Y. Weiss. ``Constructing Free Energy Approximations and Generalized Belief Propagation Algorithms,'' \emph{IEEE Trans. Info. Theory,} 2282--2312, July 2005.

\end{thebibliography}
\end{document}